\documentclass[letterpaper]{article} % DO NOT CHANGE THIS
\usepackage{aaai2026}  % DO NOT CHANGE THIS
\usepackage{times}  % DO NOT CHANGE THIS
\usepackage{helvet}  % DO NOT CHANGE THIS
\usepackage{courier}  % DO NOT CHANGE THIS
\usepackage[hyphens]{url}  % DO NOT CHANGE THIS
\usepackage{graphicx} % DO NOT CHANGE THIS
\usepackage{natbib}  % DO NOT CHANGE THIS AND DO NOT ADD ANY OPTIONS TO IT
\usepackage{caption} % DO NOT CHANGE THIS AND DO NOT ADD ANY OPTIONS TO IT
\usepackage{xcolor}
\usepackage{algorithm}
\usepackage{algorithmic}

\nocopyright % PLEASE REMOVE FOR Production LEVEL
\usepackage{newfloat}
\usepackage{listings}
\DeclareCaptionStyle{ruled}{labelfont=normalfont,labelsep=colon,strut=off} % DO NOT CHANGE THIS
\floatstyle{ruled}
\newfloat{listing}{tb}{lst}{}
\floatname{listing}{Listing}
\title{{\small\textnormal{\textit{Please read and cite our newer study, which supersedes these results: \url{https://arxiv.org/abs/2607.20300}.}}}\\[2ex]
From Hugging Face to GitHub: Tracing License Drift in the\\Open-Source AI Ecosystem}
\author{
    James Jewitt, Hao Li, Bram Adams, Gopi Krishnan Rajbahadur, Ahmed E. Hassan
}

\affiliations{
    School of Computing, Queen's University, Kingston, ON, Canada\\
    \{james.jewitt, hao.li, bram.adams\}@queensu.ca, grajbahadur@acm.org, ahmed@cs.queensu.ca

}

\usepackage{bibentry}
\usepackage[most]{tcolorbox} 
\usepackage{amsmath,amssymb,amsfonts}
\usepackage{algorithmic}
\usepackage{graphicx}
\graphicspath{{./figs/}}
\DeclareGraphicsExtensions{.png,.pdf}
\usepackage{textcomp}
\usepackage{xcolor}
\usepackage{longtable}
\usepackage{booktabs}
\usepackage{tabularx}
\usepackage{array}

\DeclareRobustCommand{\hypobox}[1]{%
\begin{tcolorbox}[  
        breakable,
        left=0pt,
        right=0pt,
        top=0pt,
        bottom=0pt,
        width=\dimexpr\columnwidth\relax, 
        enlarge left by=0mm,
        boxsep=5pt,
        arc=3pt,outer arc=3pt,
        ]
        #1
\end{tcolorbox}
}

\begin{document}

\maketitle

\newcommand{\rqone}{What licensing propagation patterns emerge when downstream software applications adopt licensing from upstream components in the open-source machine learning supply lineage chain?}
\newcommand{\rqtwo}{Given the license propagation patterns uncovered, what proportion of artifact links violate established compatibility rules, as detected by the LicenseRec framework?}

\newcommand{\motivation}{\noindent\emp{Motivation. }}
\newcommand{\approach}{\medskip\noindent\emp{Approach. }}
\newcommand{\findings}{\medskip\noindent\emp{Findings. }}

% Input the abstract
\begin{abstract}

Hidden license conflicts in the open-source AI ecosystem pose serious legal and ethical risks, exposing organizations to potential litigation and users to undisclosed risk. However, the field lacks a data-driven understanding of how frequently these conflicts occur, where they originate, and which communities are most affected. 
We present the first end-to-end audit of licenses for datasets and models on Hugging Face, as well as their downstream integration into open-source software applications, covering 364 thousand datasets, 1.6 million models, and 140 thousand GitHub projects. Our empirical analysis reveals systemic non-compliance in which 35.5\% of model-to-application transitions eliminate restrictive license clauses by relicensing under permissive terms. In addition, we prototype an extensible rule engine that encodes almost 200 SPDX and model-specific clauses for detecting license conflicts, which can solve 86.4\% of license conflicts in software applications.
To support future research, we release our dataset and the prototype engine.
Our study highlights license compliance as a critical governance challenge in open-source AI and provides both the data and tools necessary to enable automated, AI-aware compliance at scale.
\end{abstract}

% Input the sections
\section{Introduction}
\label{sec:introduction}

The open-source AI ecosystem is navigating a legal minefield, with high-stakes copyright litigation threatening the operational foundations of the industry. Recent court rulings have created a volatile and uncertain landscape. On one hand, courts in cases like Bartz v. Anthropic~\citep{bartz_v_anthropic_2024} have found AI training to be an "exceedingly transformative" fair use. On the other, the sheer cost of litigation can be catastrophic, regardless of the verdict. The case of Ross Intelligence~\citep{thomson_reuters_v_ross_2020}, which was forced to cease operations~\citet{aba_ross_shutdown_2020} under the financial strain of its legal defence long before a final judgment, serves as a stark warning of the existential risks involved. This intense legal scrutiny has, until now, focused almost exclusively on the legitimacy of training data, demonstrating a critical need for proactive compliance frameworks that can help organizations avoid such costly disputes altogether.

However, this focus on training data overlooks the full scope of legal risk embedded in the AI supply chain: the end-to-end lineage of digital artifacts from datasets, the raw materials of AI; to the models trained on them; and into the final software applications that use them. A systemic governance failure is occurring at every stage of this chain, driven by license drift: the process by which legal and ethical obligations are stripped away as artifacts propagate downstream. This creates two distinct but interconnected points of legal exposure: first, at the model creation stage, where training data licenses are often disregarded; and second, further downstream, where the model's own license is violated upon integration. While foundational studies have mapped these stages in isolation ~\citet{jiang2024,stalnaker2025}, the cumulative, end-to-end nature of this compliance failure has remained unquantified, leaving the ecosystem's true legal exposure dangerously underestimated.

This paper presents the first end-to-end empirical audit of this comprehensive compliance crisis. We quantify the staggering scale of license drift by tracing the license lineage of 364,000 datasets and 1.6 million models from Hugging Face into 140,000 GitHub applications. Our analysis reveals systemic non-compliance across the entire chain, culminating at a critical failure point: the model-to-application stage, where 35.5\% of transitions violate the upstream model's license. To combat this, we introduce LicenseRec: a novel, AI-aware framework, implemented as a proof-of-concept prototype, that automatically detects these conflicts and recommends compliant licensing solutions, demonstrating a path toward mitigating this widespread risk.

We summarize our primary contributions below:

\begin{itemize}
    \item We conduct the first large-scale, end-to-end empirical audit of license propagation in the AI supply chain, providing quantitative evidence of the systemic compliance risk that harms both creators and practitioners.
    
    \item We design and implement LicenseRec, a novel, AI-aware framework capable of automatically detecting license conflicts and recommending compliant solutions.
    
    \item We publicly release our comprehensive dataset of license lineage and the open-source LicenseRec prototype to empower future research and automated compliance.
\end{itemize}

\section{Related Work}
\label{sec:related}
\begin{figure*}[t]
    \centering
    \includegraphics[width=0.9\textwidth]{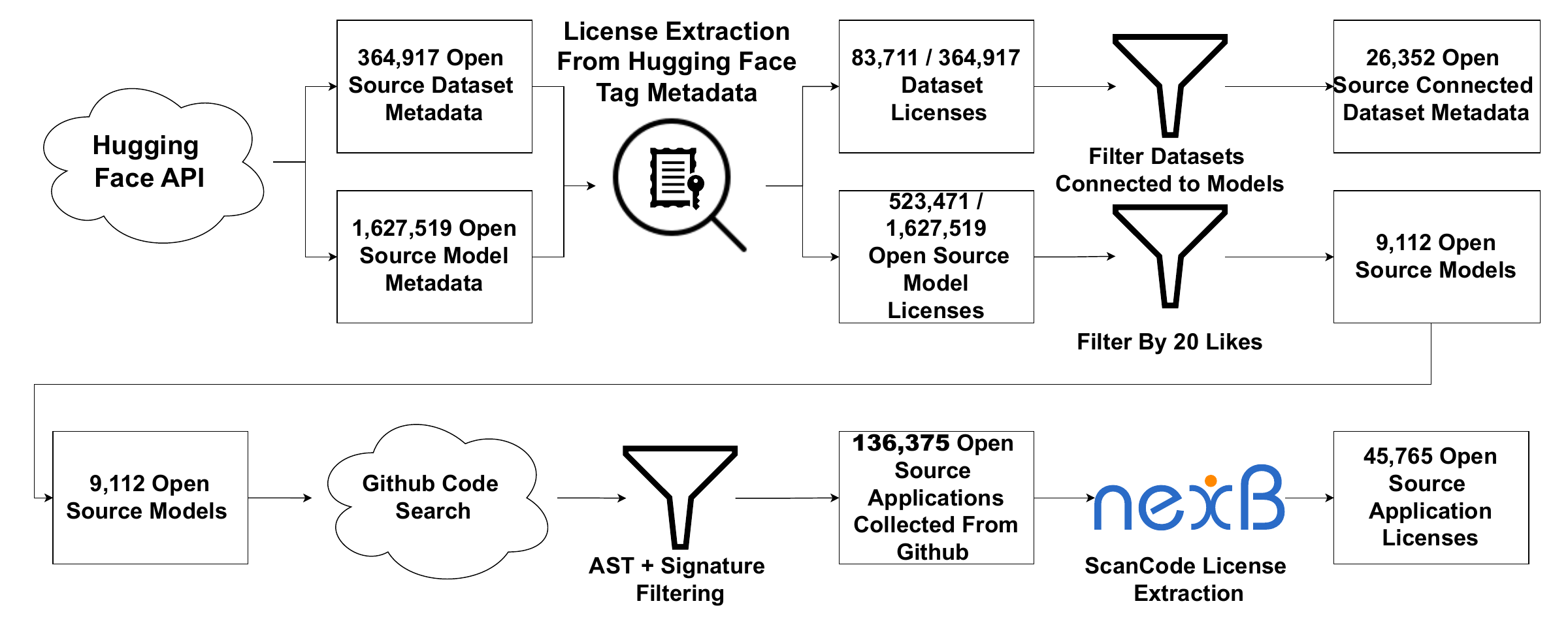}
    \caption{A comprehensive overview of our multi-stage data collection approach. }
    \label{fig:mining-process}
\end{figure*}

\paragraph{Empirical Studies of AI Supply Chains} Recent studies of the AI supply chain have revealed significant license inconsistencies within isolated segments, highlighting the need to understand how these issues propagate from end to end. The AI supply chain is the lineage of artifacts from datasets, to models, and into the final applications that use them. Foundational empirical work on this chain began with the PeaTMOSS dataset~\citep{jiang2024}, which provided the first large-scale mapping of the model $\rightarrow$ application link. This study revealed widespread inconsistencies between the licenses of pre-trained models and their downstream dependent projects. Complementing this, research by~\citet{stalnaker2025} conducted an extensive audit of the preceding dataset $\rightarrow$ model link on the Hugging Face platform. They documented significant license drift, finding that license declarations frequently change between a dataset and a model but did not analyze the compatibility implications of these changes. While these studies provided critical evidence of compliance issues at distinct stages, the cumulative effect of these problems across the full supply chain remained unknown. Our work directly addresses this gap by conducting the first end-to-end audit, connecting these segments to trace how license obligations are passed, altered, and discarded from the original dataset to the final software application.

\paragraph{The Interpretive Challenge of Software Licensing}
Auditing this supply chain is complicated by the fact that license compliance is not a simple technical check, but a complex interpretive challenge. This challenge is empirically documented by~\citet{wintersgill2024thelawdoesntwork}. Through surveys and interviews with legal experts, they found that the proliferation of licenses, the use of ambiguous legal terms, and a lack of clear court precedent create a significant "grey area," making definitive compatibility rulings difficult even for professionals. Their findings confirm the reality that, when it comes to licensing, "The Law Doesn't Work Like a Computer." 

Our goal is not to offer a definitive legal interpretation but to adopt a formal, replicable model suitable for large-scale computational analysis. To this end, we build upon the theoretical foundation provided by~\citet{moreau2019}, who deconstruct license terms into fundamental, machine-readable statuses: Permission, Duty, and Prohibition. We acknowledge this logic-based approach differs from a more risk-averse legal perspective which, as highlighted by the practitioners in the~\citet{wintersgill2024thelawdoesntwork} study, often considers cumulative complexity and practical risk over direct logical contradiction. For instance, while combining multiple licenses that each impose a Duty to display a separate notice does not create a logical contradiction, a legal practitioner might still deem the combination problematic due to the real-world usability and compliance risks involved. For our large-scale empirical audit, we adopt the more direct, contradiction-focused model as it provides a clear, replicable, and automatable basis for analysis.

\paragraph{Frameworks for Automated License Analysis}
To translate these legal principles into practice, researchers and practitioners most commonly rely on compatibility matrices. These frameworks work by pre-calculating and storing the compatibility relationships between pairs of common licenses. Prominent examples include the matrix underlying the European Commission's Licensing Assistant (JLA), which helps users select and compare licenses, and the comprehensive matrix maintained by the Open Source Automation Development Lab (OSADL). The prerequisite for using any such framework is a process of license categorization, where hundreds of unique license strings are grouped into a smaller set of canonical categories based on their core obligations. This practice is guided by the principles of organizations like the Free Software Foundation (FSF), which classifies licenses based on their adherence to user freedoms and their copyleft requirements.

However, these traditional frameworks have a critical limitation in the modern AI ecosystem. Their compatibility rules and categories, designed for conventional software, are ill-equipped to model the novel, use-based restrictions now common in AI-specific licenses. For example, a traditional matrix would not capture a prohibition on using a model for military applications or a duty to prevent a model from generating misinformation. This gap highlights the need for a new, AI-aware compliance framework that augments traditional compatibility matrices with an understanding of the unique obligations present in modern machine learning licenses.

\section{Methodology}
\label{sec:approach}

\begin{figure}[t]  
    \centering
    \includegraphics[width=\columnwidth]{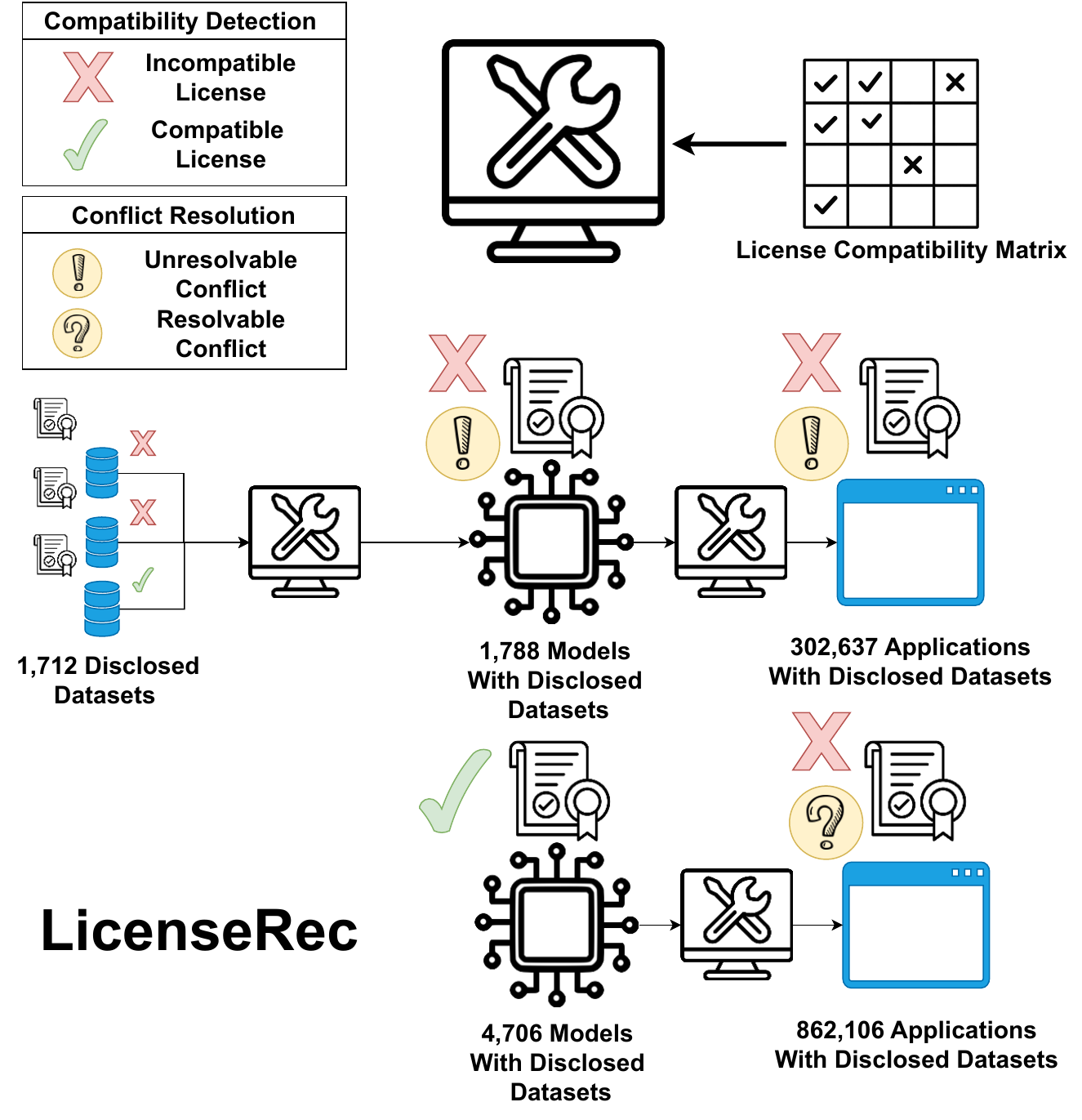}
    \caption{An overview of our LicenseRec engine.}
    \label{fig:licenserec}
\end{figure}

Our methodology comprises three stages to conduct an end-to-end audit of the open source AI supply chain, as illustrated in Figures~\ref{fig:mining-process} and ~\ref{fig:licenserec}. The first stage is Data Collection, where we construct a dependency graph linking datasets and models from Hugging Face to downstream applications on GitHub. The second stage introduces the LicenseRec, a rule-based engine we develop to automatically detect license conflicts and recommend compliant solutions. The final stage is the Evaluation, where we benchmark LicenseRec's performance against other compatibility matrices and tools. 

\subsection{Data Collection}

Our analysis requires data spanning the entire AI supply chain, from the initial datasets to the models trained on them, and finally to the software applications integrating these models. Our audit encompasses both open-source and proprietary model ecosystems. It is important to note that for proprietary models, our analysis is limited to the model $\rightarrow$ application link, as their upstream training data typically is not disclosed. 

\begin{table}[t]
\centering
\caption{Comparison of Data Sources}
\label{tab:items_comparison}
\begin{tabular}{@{}lrrr@{}}
\toprule
                     & \# Datasets & \# Models & \# Repos \\ \midrule
\citet{stalnaker2025} & 175,000     & 760,460   & \textemdash \\
\citet{jiang2024}     & \textemdash & 281,638   & 28,575    \\ \midrule
\textbf{Ours}         & \textbf{364,917}     & \textbf{1,627,519} & \textbf{136,375}   \\ \bottomrule
\end{tabular}
\end{table}

Our work performs the first large-scale, end-to-end audit of the open source AI supply chain. The scale of our dataset is substantially larger than in prior work ~\citep{stalnaker2025, jiang2024}, encompassing 364,917 datasets and 1,627,519 models from Hugging Face, roughly double the 175,000 datasets and 760,460 models analyzed by ~\citet{stalnaker2025} (see Table~\ref{tab:items_comparison}). However, our key contribution moves beyond just scale to address the scope of the supply chain. While previous studies analyzed isolated segments, our audit is the first to construct a full dataset $\rightarrow$ model $\rightarrow$ repository lineage by linking the most popular of these upstream Hugging Face artifacts to nearly 140,000 downstream GitHub software repositories. This complete linkage enables a comprehensive view of how license obligations propagate across the entire development lifecycle.

We collect metadata for all available datasets and models hosted on the Hugging Face platform. Following an adapted methodology from ~\citet{stalnaker2025}, we query the Hugging Face API using an authenticated access token. This process, conducted through April 22, 2025, yields metadata for 364,917 datasets and 1,627,519 models.

We process the collected metadata for both datasets and models to establish license information and supply chain links, primarily using the tags field. Since dataset identifiers in this field are often incomplete (e.g., missing an organization name), we normalize them by searching Hugging Face to resolve each dataset's fully qualified identifier, ensuring consistent dataset $\rightarrow$ model links. 

To trace model usage in downstream applications, we select a subset of 9,112 popular models from Hugging Face, defined as those with over 20 ``likes''. While prior work used download counts, we use ``likes'' as our popularity metric because recent research by~\citet{kadasi2025modelhubsbeyondanalyzing} shows they are a strong proxy for community adoption and, unlike download metrics, are consistently available via the Hugging Face API.

For each selected model, we use its author and ID (e.g., microsoft/codebert-base) as a search query via the GitHub Codesearch API, filtering results to include only Python source files (.py). This process yields 1,164,743 (.py) files, located in 136,375 unique repositories. To extract application licenses, we first download each repository, filter for all relevant license and notice files~\citep{Zacchiroli_2022}, and then run the ScanCode toolkit on this subset. We chose this toolkit because its ability to detect multiple licenses within a single repository provides a more holistic perspective than the single-license detection offered by the GitHub API or used in prior studies like~\citet{jiang2024}.

To confirm a model is actively used, we perform static analysis guided by code signatures. We want to ensure that the model is called in the code and not just referenced in a comment. Our verification process adapts the filtering techniques from the PeaTMOSS dataset~\citep{jiang2024}. We parse each Python file into an Abstract Syntax Tree (AST), a structural representation of the source code. The AST is crucial because it allows us to reliably distinguish active code from non-executable lines that were commented out later in the code. Within the parsed code, we identify signatures of how the model is run using a comprehensive, manually curated set of 11,000 code signatures. For replicability, we use Python's native ast library, and the full list of signatures is available in our replication package. This filtering process confirms 136,375 repository entries that actively integrate 6,999 unique open-source Hugging Face models.

\subsection{LicenseRec}
\label{subsec:licence_categorization}

Analyzing license propagation is difficult because the hundreds of unique license strings in our dataset make direct comparison complex. To address this, we develop a set of consolidated license categories to group licenses with similar rights and obligations. This is crucial for practitioners like Luis. He may not know the specific legal differences between an `MIT' and an `Apache-2.0' license, but he knows he wants his work to be freely used by anyone for any purpose. Our Permissive category allows him to see at a glance that both licenses align with his values. Conversely, if Luis wanted to ensure any modifications to his project also remain open source, he could simply look for licenses grouped in the Copyleft category. 

The full definitions for these categories are provided in Appendix~\ref{app:category}. Our categorization adapts the engine from ~\citet{stalnaker2025}, and refines it based on established principles from the Free Software Foundation (FSF)~\citep{fsf2025} and Creative Commons~\citep{cc2025}, for instance, by using the FSF's concrete categories like Permissive, Share-Alike, CopyLeft. We also choose to subdivide the CC licenses to account for their varying obligations (e.g., Non-Commercial, Share-Alike).

\begin{table}[t]
    \centering
    \small
    \setlength{\tabcolsep}{3pt}
    \caption{Distribution of License Categories Across Datasets, Models, and Repositories}
    \label{tab:license_category_distribution}
    \begingroup
    \begin{tabularx}{\columnwidth}{lrrrrrr}
        \toprule
        \textbf{Category} & \multicolumn{2}{c}{\textbf{Datasets}} & \multicolumn{2}{c}{\textbf{Models}} & \multicolumn{2}{c}{\textbf{Repos}} \\
        \cmidrule(lr){2-3} \cmidrule(lr){4-5} \cmidrule(lr){6-7}
        & \textbf{\#} & \textbf{\%} & \textbf{\#} & \textbf{\%} & \textbf{\#} & \textbf{\%} \\
        \midrule
        Permissive     & 64,088 & 70.8 & 380,022 & 62.8 & 465,053 & 90.6 \\
        CopyLeft       &  1,326 &  1.5 &   3,898 &  0.6 &  32,603 &  6.3 \\
        ML             &  8,323 &  9.2 & 114,797 & 19.0 &     995 &  0.2 \\
        SA             &  2,939 &  3.2 &   3,640 &  0.6 &   2,409 &  0.5 \\
        PD             &  1,851 &  2.0 &   2,192 &  0.4 &   3,938 &  0.8 \\
        NC             &  2,079 &  2.3 &  13,779 &  2.3 &   4,844 &  0.9 \\
        NC SA          &  1,930 &  2.1 &   3,735 &  0.6 &   2,672 &  0.5 \\
        NC ND          &    947 &  1.0 &   1,250 &  0.2 &     842 &  0.2 \\
        ND             &    228 &  0.3 &     158 &  0.0 &      41 &  0.0 \\
        UNKNOWN        &  6,758 &  7.5 &  81,896 & 13.5 &     164 &  0.0 \\
        \midrule
        \textbf{Total}      & \textbf{90,469} & \textbf{100.0} & \textbf{605,367} & \textbf{100.0} & \textbf{513,561} & \textbf{100.0} \\
        \bottomrule
    \end{tabularx}
    \endgroup
\end{table}

\begin{figure*}[t]
\centering
\includegraphics[width=0.75\textwidth]{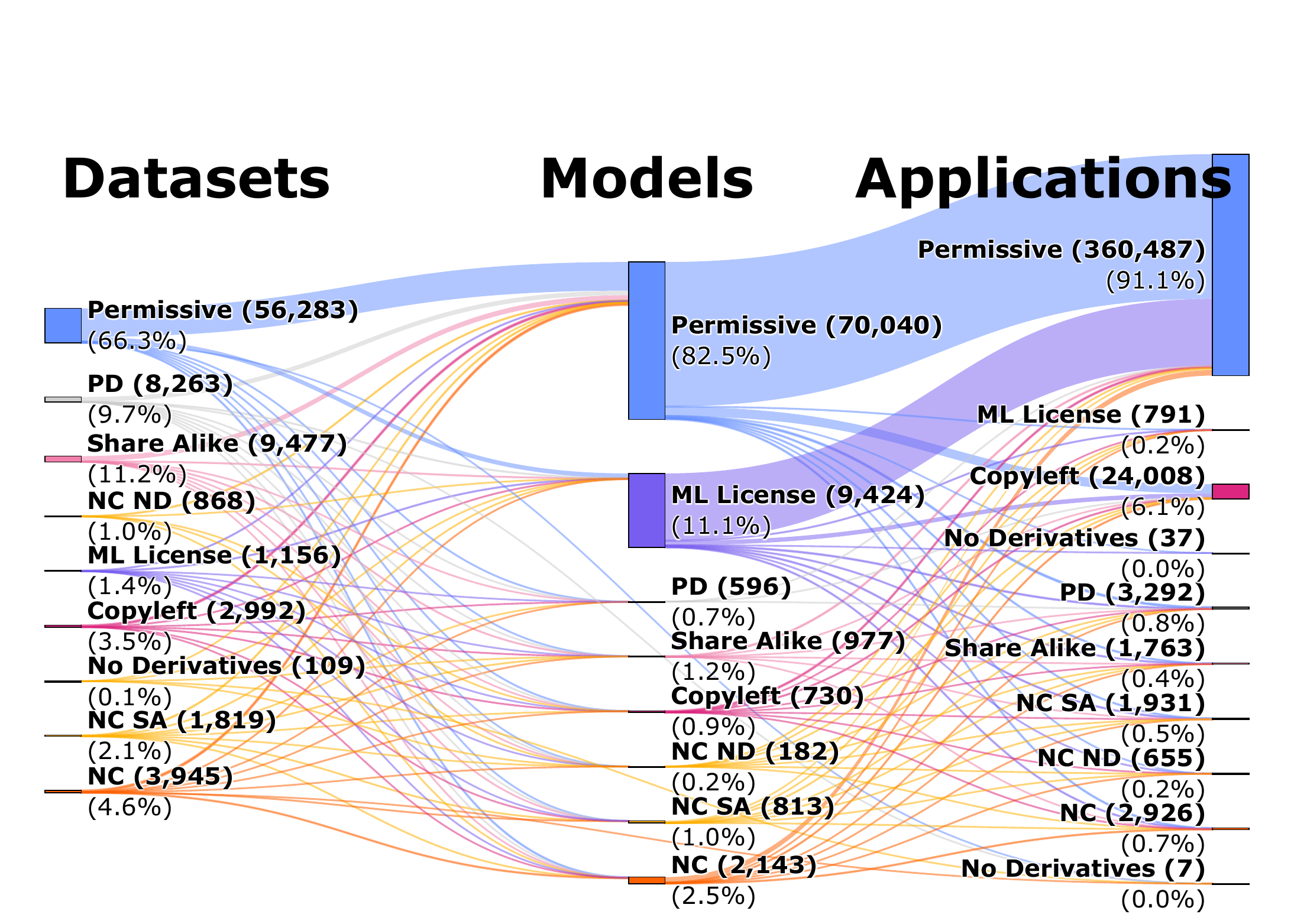}
\caption{License Category Transitions: A Sankey diagram illustrating the flow of license categories from datasets to models and finally to code repositories.}
\label{fig:license_sankey}
\end{figure*}

The resulting categories serve a dual purpose in our methodology. First, they enable our empirical analysis of license propagation. The distribution of these license can be found in Table~\ref{tab:license_category_distribution}. By mapping each raw license to a category, we can track how licensing patterns change as the artifacts move from upstream to downstream in the open-source supply chain. Second, these categories form the user-facing core of our LicenseRec engine. When assessing compatibility, LicenseRec recommends a set of valid licenses, based on their categorized level of permissiveness. 

We empirically analyze the license transitions across this graph by mapping the licenses for all dataset $\rightarrow$ model, model $\rightarrow$ application and end-to-end dataset $\rightarrow$ application, links to our predefined license categories in Table~\ref{tab:license_categories}. With these categorized links, we then perform a frequency analysis to identify common propagation patterns. This involves quantifying how often the specific license pathways occur (e.g., public domain $\rightarrow$ permissive). This analysis aims to reveal the common licensing strategies developers employ, as well as how often license obligations are respected. 

To formally analyze license interactions, we adopt the framework from~\citet{moreau2019} which deconstructs license terms into fundamental statuses for any action: Permission, Duty (an obligation to perform an action), and Prohibition (a ban on an action). Within this model, we define license compatibility as a state where the combined duties and prohibitions from two licenses are not self-contradictory. A license conflict occurs when one license imposes a Duty that another Prohibits. For instance, a conflict arises if an upstream model’s license includes a duty to ‘share source code upon distribution,’ while the downstream application’s license prohibits it.

To automate conflict detection, we developed a comprehensive, machine-readable compatibility matrix. Our process began by adopting the well-established Open Source Automation Development Lab (OSADL) matrix~\citep{osadl2025} for traditional licenses. We then extended it in two ways: first, by encoding the official compatibility rules published by Creative Commons~\citep{cc2025}, and second, by manually analyzing the unique use-based restrictions and redistribution duties common in modern ML licenses. Notably, this analysis revealed a common requirement for ML licenses: downstream applications must include the original license terms to ensure ethical and use-based restrictions are preserved, even if the application's own code is under a different license.

For conflict detection, LicenseRec takes licenses of any two linked upstream and downstream artifacts in a supply chain and reports whether they are compatible according to the matrix. For license recommendations, the engine takes a downstream artifact (e.g., a new application) and the full set of its upstream component licenses as input. If the downstream license is not compatible, LicenseRec intersects the compatibility sets of every upstream license to get a subset that would satisfy all upstream obligations. If any fundamental, unresolvable conflicts exist, it stops here. It then filters this subset with a configurable whitelist, discards older Creative Commons licenses when a 4.0 variant is available, and ranks the options based on real-world frequency of licenses in their projects and  returns up to five licenses per SPDX category.  

We assess the Fixability of these violations. For every identified conflict, we use LicenseRec's recommendation function to determine if a set of valid, compatible downstream licenses exists. We measure Fixability as the percentage of initial violations for which LicenseRec could recommend at least one valid, compatible license category for the downstream artifact. This quantifies how many existing violations are due to sub-optimal license choices versus how many are fundamental incompatibilities.

\textbf{Evaluation.}
We benchmark the effectiveness of LicenseRec's logic. We compare the compatibility rulings of LicenseRec, which is based on our augmented matrix, against the outcomes produced by the compatibility matrix used in the PeaTMOSS study~\citep{jiang2024}. We compare our Fixability compared to the EU's Licensing Assistant tool~\citep{licensing_assistant2025}. This two-part benchmark contextualizes the performance and coverage of our framework.

\section{Results}
\label{sec:results}

\subsection{License Drift in AI Ecosystem}

\textbf{Nearly all license categories exhibit disregard with their obligations, frequently being replaced by permissive licenses in downstream components.} This trend is evident in Figure~\ref{fig:license_sankey}, which reveals exceptionally low retention rates for most non-permissive categories. For instance, Non-Commercial licenses, which impose significant use restrictions, have a retention rate of only 20.7\%. Similarly, only 3.9\% of Share-Alike datasets resulted in a model with the same license category, effectively ignoring the licenses primary copyleft obligation. 

\textbf{This trend of disregarding restrictive obligations culminates at the model $\rightarrow$ repository stage, which acts as a filter that erases nearly all remaining upstream license obligations.} This is detailed in Figure~\ref{fig:license_sankey}, this stage is marked by a near-total collapse in the retention rates for non-permissive licenses. Strikingly, ML License category, has specific use-based restrictions have a retention rate of just 0.4\%. This collapse in obligation retention is mirrored by other restrictive categories like Non-Commercial (6.7\%) and and Share-Alike (1.7\%). As visualized in Figure~\ref{fig:license_sankey}, these flow almost entirely into the PERMISSIVE category, which accounts for 91.1\% of all repository licenses in these chains. This finding is a critical disconnect; application developers appear to treat complex restrictive licensed models as simple permissive libraries, ignoring their terms at the final point of integration. 

\textbf{Permissive licenses exhibit high stability as they move downstream.} This pattern is quantified in Figure~\ref{fig:license_sankey}, which shows that in the dataset $\rightarrow$ model stage, 82.8\% of models retained a permissive license, a rate that increased to 91.9\% in the subsequent model $\rightarrow$ repository stage. This consistent Permissive $\rightarrow$ Permissive flow is visualized by Figure~\ref{fig:license_sankey}, and establishes this category as the most stable licensing choice throughout the supply chain.

\textbf{Despite the trend towards permissive licensing, a notable trend emerges at the final model $\rightarrow$ repository stage.} As Figure~\ref{fig:license_sankey} shows, while the retention rates for most restrictive licenses collapse at this stage - such as ML License (0.4\%) and Share-Alike (1.7\%) - the retention rate for COPYLEFT models is substantially higher at 25.3\%. This suggests that while model creators tend to avoid copyleft, a larger group of application developers actively chooses it for their final product.

\hypobox{
\textbf{Summary:} 
Our audit reveals a systematic pattern of non-compliance where restrictive license obligations are progressively discarded. This trend culminates at the critical model $\rightarrow$ repository stage, where compliance collapses almost completely; a mere 0.4\% of repositories retain the obligations of the ML Licenses used by their models. This creates a significant and systemic compliance risk. The one notable exception is the Copyleft category, which shows a substantially higher retention rate of 25.3\% in the final application stage, suggesting a resilient subset of developers committed to reciprocal sharing.

}
% \subsection*{\rqtwo \todo{remove this subsection or shorten it}}

\begin{figure}[t]
\centering
\includegraphics[width=\columnwidth]{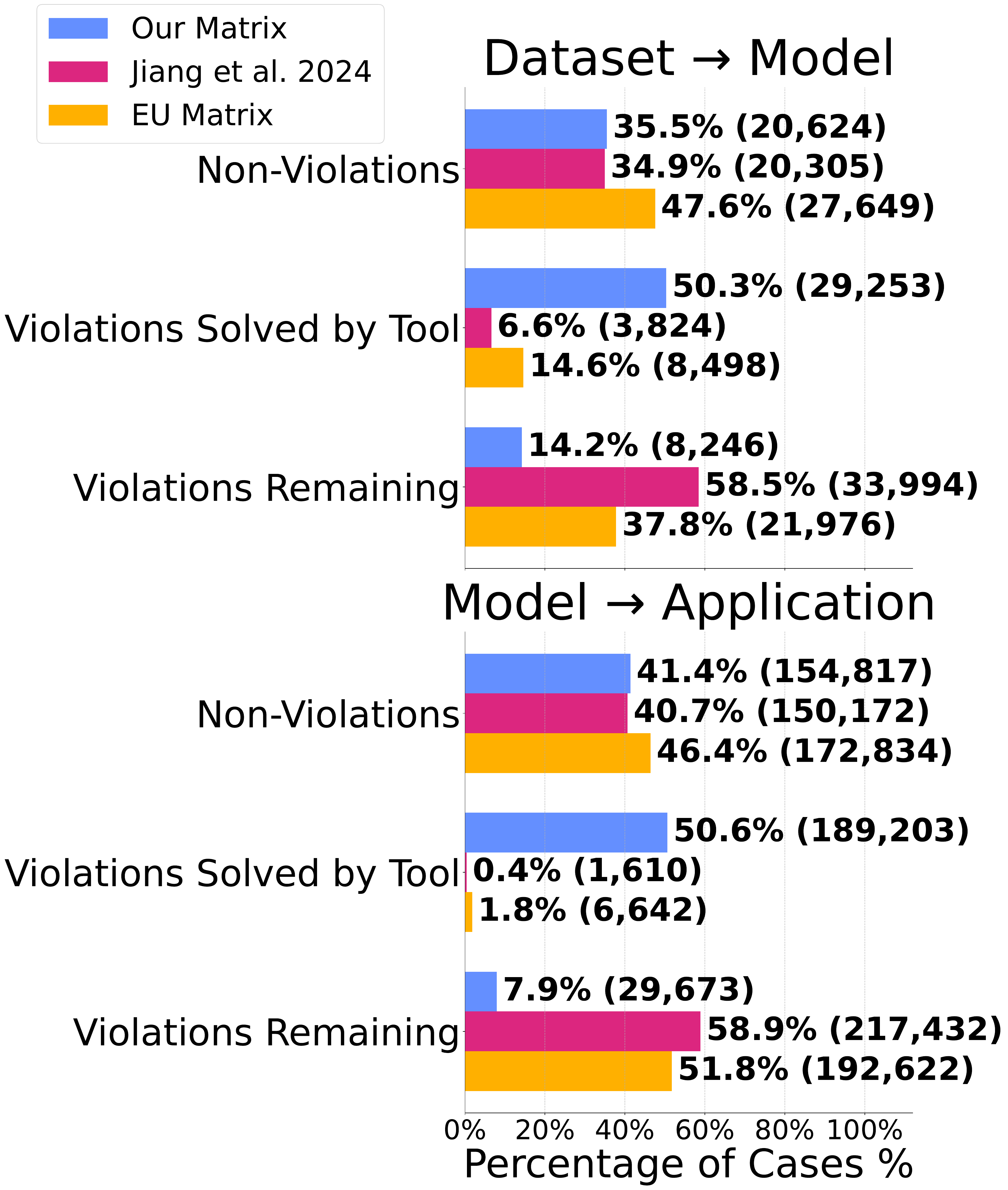}
\caption{Summary of license violation analysis. Top: dataset $\rightarrow$ model violations for models with linked datasets. Bottom: model $\rightarrow$ repository violations.}
\label{fig:analysis_figure_13}
\end{figure}

\subsection{Effectiveness of LicenseRec}

\begin{table*}[t]
\centering
\caption{Top Most Frequent License Violation Patterns by Stage, calculated for links where both upstream and downstream licenses were successfully categorized and known in our matrix}
\label{tab:violation_patterns}
\resizebox{\textwidth}{!}{%
\begin{tabular}{@{} l l r r | l r r | l r r @{}}
\toprule
\multicolumn{4}{c}{\textbf{Dataset $\rightarrow$ Model}} & \multicolumn{3}{c}{\textbf{Model $\rightarrow$ Repository}} & \multicolumn{3}{c}{\textbf{Dataset $\rightarrow$ Repository}} \\
\cmidrule(r){1-4} \cmidrule(l){5-7} \cmidrule(l){8-10}
\textbf{Rank} & \textbf{Violation Pattern} & \textbf{Count} & \textbf{\%} & \textbf{Violation Pattern} & \textbf{Count} & \textbf{\%} & \textbf{Violation Pattern} & \textbf{Count} & \textbf{\%} \\
\midrule
1 & SA $\rightarrow$ P & 5,208 & 37.4\% & ML $\rightarrow$ P & 109,214 & 84.9\% & SA $\rightarrow$ P & 32,312 & 63.4\% \\
2 & NC $\rightarrow$ P & 2,620 & 18.8\% & NC $\rightarrow$ P & 8,980 & 7.0\% & CL $\rightarrow$ P & 8,184 & 16.1\% \\
3 & CL $\rightarrow$ P & 2,372 & 17.0\% & NC-SA $\rightarrow$ P & 3,338 & 2.6\% & NC $\rightarrow$ P & 6,442 & 12.6\% \\
4 & NC-SA $\rightarrow$ P & 992 & 7.1\% & P $\rightarrow$ PD & 2,117 & 1.6\% & P $\rightarrow$ PD & 1,952 & 3.8\% \\
5 & NC-ND $\rightarrow$ P & 775 & 5.6\% & CL $\rightarrow$ P & 1,702 & 1.3\% & NC-SA $\rightarrow$ P & 997 & 2.0\% \\
6 & ML $\rightarrow$ P & 560 & 4.0\% & SA $\rightarrow$ P & 1,323 & 1.0\% & SA $\rightarrow$ PD & 382 & 0.7\% \\
7 & P $\rightarrow$ PD & 336 & 2.4\% & ML $\rightarrow$ PD & 1,060 & 0.8\% & ML $\rightarrow$ P & 334 & 0.7\% \\
8 & NC $\rightarrow$ ML & 227 & 1.6\% & ML $\rightarrow$ SA & 427 & 0.3\% & NC-ND $\rightarrow$ P & 224 & 0.4\% \\
9 & NC-SA $\rightarrow$ ML & 201 & 1.4\% & NC-ND $\rightarrow$ P & 268 & 0.2\% & CL $\rightarrow$ PD & 70 & 0.1\% \\
10 & NC-SA $\rightarrow$ SA & 173 & 1.2\% & NC $\rightarrow$ PD & 60 & 0.0\% & NC $\rightarrow$ PD & 27 & 0.1\% \\
\midrule[\heavyrulewidth]
\textbf{Total Violations} & \multicolumn{2}{r}{\textbf{13,921}} &  & \multicolumn{2}{r}{\textbf{128,568}} &  & \multicolumn{2}{r}{\textbf{50,977}} &  \\
\textbf{Violation Rate} & \multicolumn{2}{r}{\textbf{(13,921/79,667) 17.5\%}} &  & \multicolumn{2}{r}{\textbf{(128,568/362,424) 35.5\%}} &  & \multicolumn{2}{r}{\textbf{(50,977/204,307) 25.0\%}} &  \\
\bottomrule
\end{tabular}%
}
\end{table*}

\textbf{Our Analysis shows that LicenseRec successfully fixes a majority of identified license violations, correcting 78.0\% of the conflicts at the initial dataset $\rightarrow$ model stage, and an even greater 86.4\% at the final model $\rightarrow$ repository stage.} The summary of this evaluation, presented in Figure~\ref{fig:analysis_figure_13}, indicates that many licenses issues in the ecosystem are not due to irreconcilable issues. Instead they appear be the result of suboptimal license selection - a problem that can be systematically identified and fixed with automated framework. 

\textbf{A small number of high-risk violation patterns account for the vast majority of license conflicts throughout the supply chain.}  Table~\ref{tab:violation_patterns} reveals that these violations are not random but follow predictable pathways where restrictive obligations are systematically dropped. This failure to adhere to specific license obligations is immediately evident; for example, the Share alike $\rightarrow$ permissive pattern accounts for 37.4\% of all conflicts at the initial dataset $\rightarrow$ model stage. The problem becomes more acute with ML specific licenses whose use base restrictions are frequently erased at the final, most critical stage of integration. This is demonstrated by the ML license $\rightarrow$ Permissive transition, which alone accounts for an overwhelming 84.9\% of all model $\rightarrow$ repository violations.  Compounding this is a consistent and legally precarious pattern of using non-commercial assets in permissive projects, as the non-commercial $\rightarrow$ Permissive violation ranks as a top-three conflict at every stage of analysis. These concentrated failure patterns demonstrate a systemic blind spot in the community, where developers repeatedly default to permissive licenses, ignoring the specific and varied obligations of their upstream dependencies.

\textbf{Despite the framework's effectiveness, a persistent core of conflicts remains unresolvable, indicating fundamental incompatibilities inherited from upstream components.} As shown in Figure~\ref{fig:analysis_figure_13}, 14.2\% of the violations at the initial dataset $\rightarrow$ model stage could not be resolved by simply changing the downstream license. These unresolvable conflicts occur when no valid license category can satisfy all upstream obligations. For example, a conflict arising from a model trained on a dataset with a non-commercial restriction cannot be fixed by re-licensing the model; the incompatibility is baked into the model itself. Consequently, any application using this model inherits the same unresolvable conflict. The only solution is for the developer to select a different upstream model. This underscores the limits of automated license selection; while a framework can correct errors in license declaration, it cannot fix flawed component choices, reinforcing that developer diligence in vetting the entire supply chain remains critical.

\hypobox{
\textbf{Summary:} Our investigation reveals that license conflicts are not random errors, but are dominated by specific, high-risk patterns like the systematic erasure of ML and non-commercial license obligations. We demonstrate that while a majority of these are correctable declaration errors, a significant number are fundamental incompatibilities inherited from upstream components, which cannot be fixed by simple re-licensing. This finding underscores a critical dual path to compliance: an automated framework can resolve common errors, but developer diligence in vetting the entire supply chain remains indispensable to avoid baked-in, unresolvable conflicts.
}

\section{Discussion}
\label{sec:Discussion}

\begin{table}[tb]
\centering
\small
\caption{Comparison of license violation percentages (\%) across matrices. Analysis stages are dataset to model (D $\rightarrow$\ M), model to repository (M $\rightarrow$\ R), and dataset to repository (D $\rightarrow$\ R). }
\label{tab:violation_rate_comparison}
\begin{tabular}{@{} lrrr @{}}
\toprule
\textbf{Analysis Stage} & \textbf{Ours} & \textbf{Jiang et al. Matrix} & \textbf{EU Matrix} \\
\midrule
D $\rightarrow$\ M & \textbf{17.5} & 1.4 & 12.1 \\
M $\rightarrow$\ R & \textbf{35.5} & 1.1 & 3.2 \\
D $\rightarrow$\ R & \textbf{25.0} & 0.7 & 8.6 \\
\bottomrule
\end{tabular}
\end{table}

A core aspect of LicenseRec is its compatibility-based approach, which navigates a complex and often ambiguous landscape. Determining license compatibility is rarely a definitive ``yes or no'' question as discussed by~\citet{wintersgill2024thelawdoesntwork}; it is a matter of legal interpretation and risk tolerance. LicenseRec adopts a more liberal approach, similar to the EU's Licensing Assistant tool, which focuses on direct and explicit contradictions between license terms. It is important to note, however, that a stricter, more risk-averse legal analysis would likely identify a greater number of violations, particularly in cases involving multiple licenses where the combined obligations create legal complexity rather than a direct conflict.

LicenseRec's accuracy is fundamentally dependent on the quality of the input data. If a developer on Hugging Face assigns the wrong license tag, or if ScanCode fails to detect a license in a repository, the tool's analysis will be incorrect. It cannot fix errors in the source metadata.
 
Our analysis uncovers a significant trend of \textbf{a ``Gravitational Pull'' towards permissive licensing} across the open-source AI ecosystem, revealing a systematic disregard for upstream license obligations. This trend suggests developers frequently default to permissive licensing due to a combination of convenience, limited awareness of licensing intricacies, and perhaps platform defaults that favour simplicity over compliance. The cultural norms within the open-source community, which prioritize ease of collaboration and minimal friction, also contribute to this licensing behaviour, inadvertently elevating the risk of legal non-compliance. 

\textbf{ Our ML-aware compatibility matrix provides a more accurate assessment of license conflicts, reducing the detected violation rate at the critical model-to-repository stage compared to a baseline matrix.} This improvement, detailed in Table~\ref{tab:violation_rate_comparison}, highlights the inadequacy of traditional compatibility matrices for the modern AI ecosystem. Our matrix, which accounts for ML-specific license families, flags a high violation rate of 35.5\% for model $\rightarrow$ repository links. In stark contrast, the EU matrix, misses out this important information, showing a violation rate of just 3.2\%. This addition is essential for building a useful compliance tool, as an approach that does not understand the modern AI ecosystem could mislead developers into choosing a common but non-compliant license.

We identified the transition from the \textbf{model to repository as a crucial juncture where almost all non-permissive obligations are stripped away}. This high rate reveals a significant gap in developer understanding of license obligations at the integration stage. Developers may mistakenly treat models as if they were simple software libraries, overlooking the specific licensing terms of the upstream artifacts. This finding highlights the urgent need for better communication and tools for licensing at the time of integration. Future work could build upon our AST-based approach by incorporating more sophisticated static and dynamic analysis techniques to further enhance model usage  accuracy.

Our results distinguish between two primary types of license non-compliance: \textbf{fixable and unresolvable conflicts}. Many license conflicts detected by LicenseRec are fixable, highlighting that a substantial proportion of non compliance arises from suboptimal license choices rather than fundamental incompatibilities. However, our findings also expose a core of unresolvable conflicts inherited from upstream components, such as non-commercial licenses, that are fundamentally incompatible with downstream permissive usage. This distinction emphasizes that while an automated framework can significantly improve compliance, developer diligence in selecting upstream dependencies remains indispensable. Tooling alone cannot fully eliminate baked-in license conflicts. 

\subsubsection{Future Work}\label{subsubsec:Future_Work}

The next frontier for this investigation is to explore whether similar dynamics hold true in closed source contexts. Our preliminary findings are that 81.5\% of repositories using proprietary API-based services also adopt permissive licenses raises questions for future work. How do developers reconcile the obligations of a proprietary service's Terms of Use with the freedoms of their chosen open-source license? Investigating this dynamic is a vital next step to understand compliance in the rapidly growing API-driven AI ecosystem.

\section{Conclusion}
\label{sec:conclusion}

In this work, we presented the first large-scale, end-to-end empirical audit of license compliance in the open-source AI supply chain, introducing the LicenseRec engine to automatically detect and resolve conflicts. Our findings reveal a critical, dual path to compliance: while automated tooling can successfully correct a majority of licensing errors. This underscores that many are fixable instances of suboptimal choice, but the persistence of fundamental, ``unresolvable'' conflicts inherited from upstream components proves that tooling alone is insufficient. Developer diligence in vetting the entire supply chain remains indispensable. Ultimately, by providing both the empirical data to quantify this systemic risk and a practical framework to begin mitigating it, our research offers a foundational step toward a more responsible, transparent, and legally robust AI ecosystem.

\bibliography{aaai2026.bib}

\appendix
\setcounter{secnumdepth}{1}
\section{License Categorization}\label{app:category}

% \todo{add text here}
Table \ref{tab:license_categories} provides the detailed definitions for the license categories used throughout our analysis. Our approach to creating these categories is described in the LicenseRec Framework subsection of the Methodology.

\begin{table*}[t] 
\centering
\caption{Definitions of License Categories} 
\label{tab:license_categories} 

\begin{tabular}{p{2.5cm}p{11cm}p{2.5cm}}
\toprule
\textbf{Category} & \textbf{Description} & \textbf{Change From Supply Chain 2.0 ~\cite{stalnaker2025}} \\
\midrule

NC\_ND (Non-Commercial No Derivatives) & Permits non-commercial sharing but prohibits modifications and commercial use (for example, CC BY-NC-ND). & Originated from CC class \\

NC\_SA (Non-Commercial Share Alike) & Permits non-commercial modifications and sharing only if derivatives are shared under the same or compatible terms (for example, CC BY-NC-SA). & Originated from CC class \\

NC (NON-COMMERCIAL) & Restricts commercial use but may allow derivatives under varying terms (for example, CC BY-NC, other custom non-commercial licenses). & Originated from CC class\\

ND (NO DERIVATIVES) & Allows commercial use but prohibits modifications (for example, CC BY-ND). & Originated from CC class\\

COPYLEFT & Strong copyleft licenses requiring derivatives to be licensed under the same or compatible terms. & Originated from Open Source and Data Class\\

ML\_LICENSE &  Machine Learning Licenses often including specific considerations for how providers use models, as well as obligations on downstream applications similar to copyleft.  & Originated from Machine Learning Class\\

SHARE\_ALIKE & Weak copyleft licenses requiring derivatives or modifications shared under the same or compatible terms but potentially allowing linking with differently licensed code (for example, LGPL, MPL, CC BY-SA). & Originated from Open Source and Data Class \\

PERMISSIVE & Minimal restriction licenses allowing use, modification, and distribution (commercial or non-commercial)  whose primary obligation is attribution (for example, CC BY, MIT, Apache-2.0, BSD). & Originated from Open Source and Data Class\\

PUBLIC DOMAIN & No restrictions (for example, CC0, Unlicense). & Originated from Open Source Data Class\\
\bottomrule
\end{tabular}
\end{table*}

\end{document}